\documentclass{iopconfser}

\usepackage{graphicx}
\usepackage{dcolumn}
\usepackage{amsmath}
\usepackage{amssymb}
\usepackage{color}
\usepackage{multirow}
\usepackage{url}
\usepackage{booktabs}
\usepackage{tabularx} 

\usepackage[utf8]{inputenc}
\usepackage{siunitx} 
\usepackage{verbatim} 
\usepackage[normalem]{ulem}
\usepackage{placeins} 
\usepackage{float} 
\usepackage[hidelinks]{hyperref} 
\hypersetup{
colorlinks=true,    
linkcolor=blue,
citecolor=blue,
urlcolor=blue,
pdfborder={0 0 0},
bookmarksopen=true,
bookmarksopenlevel=2,
pdfpagemode=UseOutlines,
pdfstartview={Fit},
pdftitle={},
pdfauthor={Z. Salman},
pdfsubject={Advanced muon-spin spectroscopy in magnetic field},
pdfkeywords={},
pdfhighlight= /I
}

\newcommand\hly{\bgroup\markoverwith
  {\textcolor{yellow}{\rule[-.5ex]{.1pt}{2.5ex}}}\ULon}

\newcommand{\mSR}{\textmu SR}

\newcommand{\refsubfig}[2]{\hyperref[#1]{\ref*{#1}#2}}

\begin{document}
\title {The effect of magnetic fields on vertex reconstructed muon-spin spectroscopy}

\author{Pascal Isenring$^{1,2}$ Zaher Salman$^1$}
\affil{$^1$PSI Center for Neutron and Muon Sciences, 5232 Villigen PSI, Switzerland}
\affil{$^2$Physik Institut, University of Zürich, Winterthurerstrasse 190, CH-8057 Zürich}
\email{zaher.salman@psi.ch}

\begin{abstract}
The use of a Si pixel-based particle tracking scheme in \mSR\ will, among others, allow measurements using a ten-fold increased stopped muons rate and samples ten times smaller than currently possible. Here we present simulation results to assess the effects of magnetic fields on two spectrometer configurations using a two-layered tracking scheme for the incoming and outgoing particles. At a low magnetic field of up to \SI{\sim50}{mT}, the tracking and reconstruction accuracy is only minimally influenced. Beyond a magnetic field of \SI{\sim80}{mT} the tracking capabilities diminish significantly. Operating a two-layered scheme using small magnetic fields hence does not require adaptations. Only at large magnetic fields, a tracking scheme that makes use of an accurate field map or the use of at least three layers must be employed to achieve reliable particle tracking.
\end{abstract}


\section{Introduction}
Current \mSR\ experiments at continuous muon sources are limited to a stopped muons rate of \SI{\sim 40}{kHz}. This limitation arises as a direct consequence of the necessity to be able to match every incoming muon to its decay positron, for a typical data gate window of \SI{\sim 10}{\micro s}. Furthermore, \mSR\ experiments require a sample area of at least $\sim4\times\SI{4}{mm^2}$, therefore excluding materials, such as novel quantum materials, which cannot yet be produced in large quantities. In order to overcome these limitations, one can use tracking schemes for the incoming muon and emitted positrons to match each muon to its decay positron by their decay vertex. For this purpose, a novel silicon pixel-based \mSR\ spectrometer is developed at the Paul Scherrer Institute (PSI) in Switzerland. To track particles, this spectrometer uses thin high-voltage monolithic active pixel Sensors (HV-MAPS) with a spatial resolution \SI{23}{\micro m} and timing resolution better than \SI{\sim 15}{ns} \cite{heikoPhd,mandok2025arxiv}. Taking full advantage of these developments allows (i) an increase in the stopped muons rate by at least a factor of $10$ to $\sim\SI{400}{kHz}$, and (ii) a reduction of the minimal sample size by at least a factor of $10$ to a size of $1.6\times\SI{1.6}{mm^2}$. Further advantages include the ability to measure multiple samples simultaneously and the ability to limit measurements to specified areas or volumes during processing, effectively allowing the exclusion of contributions from muons landing in the sample holder or cryostat tails. 

The application of external magnetic fields during a typical \mSR\ experiment is crucial, e.g. to distinguish between local static and dynamic fields in the sample. Here we explore the general feasibility of using a two-layered particle tracking scheme to reliably reconstruct particle trajectories in a \mSR\ measurement. We find that while the track reconstruction is sufficiently accurate for applied fields below \SI{\sim 30}{mT}, higher magnetic fields will require the use of three tracking layers to obtain the required accuracy or the reliance on an accurate field map to reconstruct particle trajectories.

\section{Simulations} \label{Simulations}
\begin{figure}[h]
    \centering
    \includegraphics[width=1.0\linewidth]{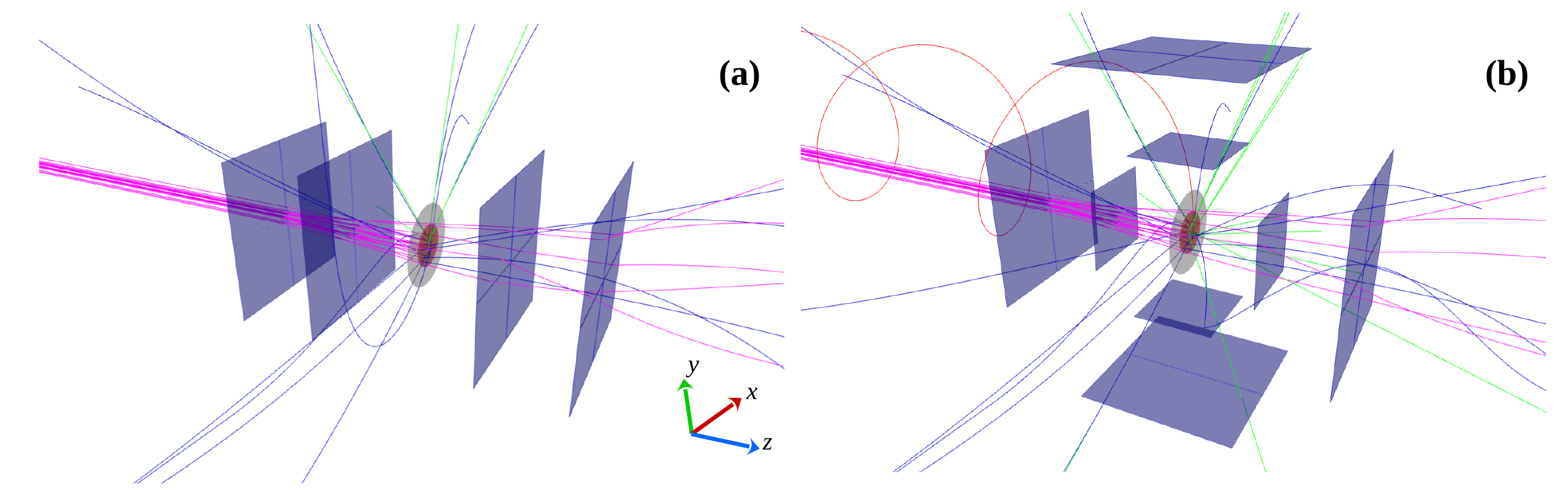}
    \caption{Depiction of the simulated \mSR\ spectrometer configurations. (a) The currently used setup \cite{mandok2025arxiv}, consisting of four layers of Si detectors, each comprised of four Si pixel chips and arranged in an upstream and downstream pairs. (b) A potential future assembly, consisting of four pairs of Si detectors layers and arranged in upstream, downstream, above and below the sample. The outer layers are comprised of four chips, whereas the inner layers are comprised of one chip each. The sample (red disk) sits between the upstream and downstream detector pairs. The gray disk serves as a non interacting (void) reference detector. Typical particle tracks are shown in an applied field of \SI{0.75}{T}, with $\mu^+$ in magenta, $e^+$ in blue, $e^-$ in red and neutrinos in green.}
    \label{spect}
\end{figure}
In order to explore the effect of magnetic fields on the tracking capabilities of a silicon pixel detector based \mSR\ spectrometer, using only two layers to track incoming and outgoing particles, we employ the musrSim simulation framework \cite{musr-sim, musr-sim2}, which is based on the GEANT4 toolkit \cite{geant4} and is tailored to the needs of the \mSR\ technique. The output of musrSim includes an event-by-event account of all incoming muons and hence allows for a straightforward examination of their tracks and those of their decay positrons. \mSR\ experiments are typically performed at muon kinetic energies of \SI{\sim4.1}{MeV}, corresponding to muon momenta of \SI{\sim28}{MeV/c} (i.e. surface muons). In the absence of external magnetic fields, the primary contribution to the uncertainty in a linear tracking scheme of particles stems from multiple Coulomb scattering within the silicon layers. The simulated \mSR\ spectrometers, shown in Fig.~\ref{spect}, consist of multiple sets of two layers. We model here two different configurations, (i) a linear configuration with two sets of identical layers, each consisting of four silicon chips arranged with one set upstream and the other downstream of the sample [Fig.~\ref{spect}(a)] and (ii) a cuboidal configuration with four sets of two layers arranged upstream, downstream, above and below the sample, the inner consisting of a single silicon chip, while the outer consists of four [Fig.~\ref{spect}(b)].
\begin{figure}[h]    \centering\includegraphics[width=0.5\linewidth]{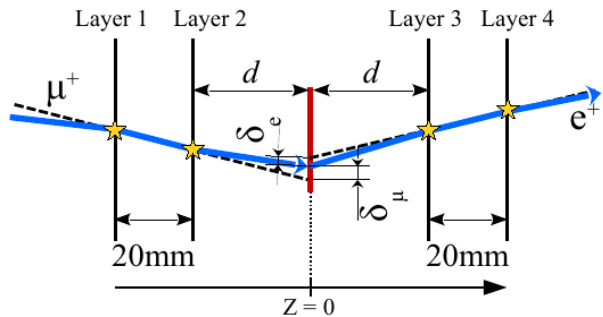}
    \caption{Schematic depiction of the tracking of incoming muons and corresponding decay positrons. Tracks consist of two hits and are extrapolated to the sample position.}
    \label{track_matching}
\end{figure}
In both configurations, the chips in each layer are glued to a \SI{25}{\micro m} polyimide foil, as is usually done when constructing Si-pixel detectors \cite{mandok2025arxiv}. The thickness of all Si chips is set to \SI{100}{\micro m}, except for the second layer with respect to the incoming muon beam, having a thickness of \SI{75}{\micro m}. This was chosen to reduce the multiple scattering of incoming muons just before entering the sample, thus reducing uncertainty in its tracking.

\begin{figure}[htb]
    \centering
    \begin{minipage}[t]{0.45\textwidth}
    \centering
        \includegraphics[width=\linewidth]{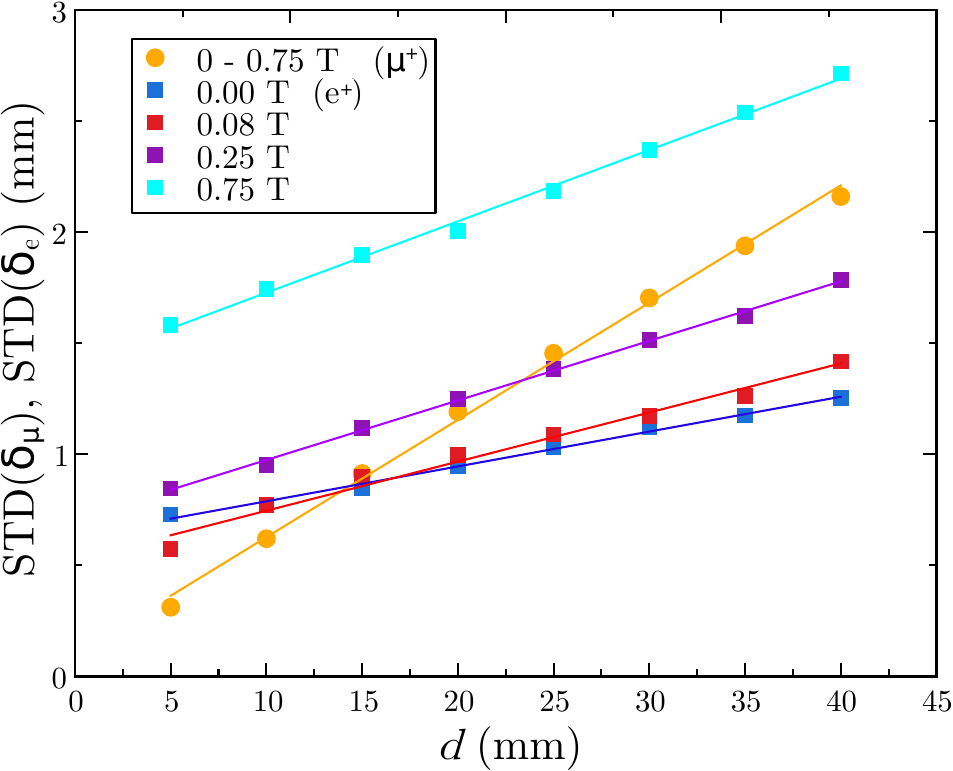}
        \caption{The STD of $\delta_\mu$ (circles) and $\delta_e$ (squares) in the linear configuration as a function of the separation between the sample and the inner detector layers for different applied magnetic fields. The STD of $\delta_\mu$ changes by less than 2\% in the range of \SIrange{0}{0.75}{T}.}
        \label{linear_diff_d_diff_B}
    \end{minipage}\hfill
    \begin{minipage}[t]{0.45\textwidth}
        \centering\includegraphics[width=\linewidth]{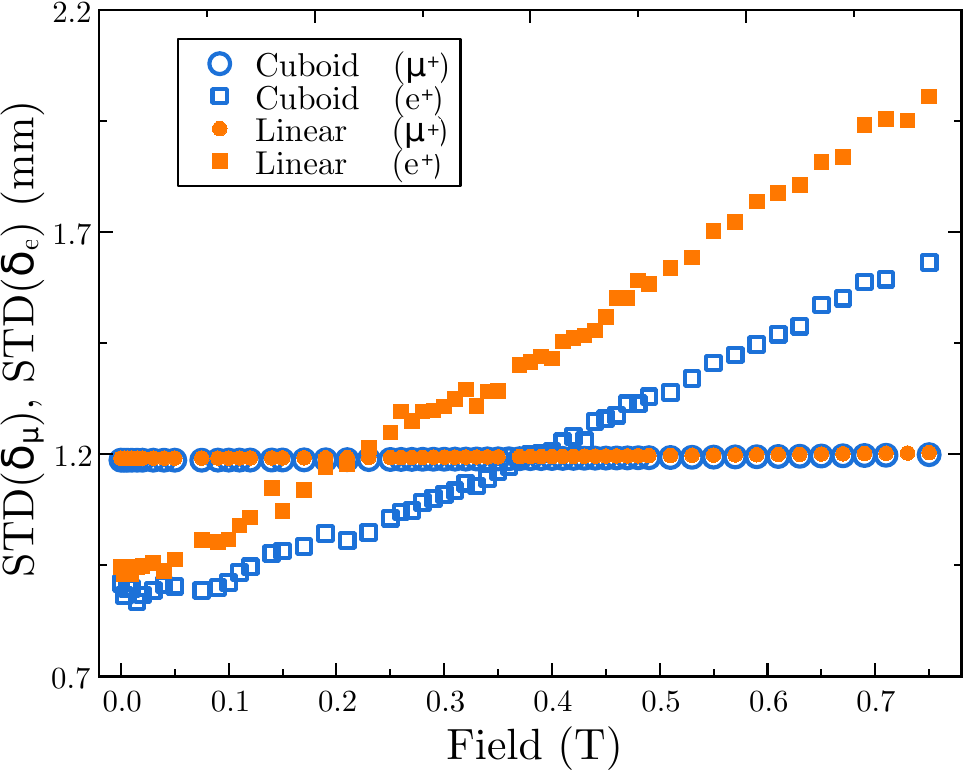}
        \caption{The STD of $\delta_\mu$ (circles) and $\delta_e$ (squares) in the linear (orange) and the cuboidal (blue) configurations as a function of applied magnetic field for a fixed distance $d=20$~mm between the sample and the inner layers.}
        \label{b_vs_sigma}
    \end{minipage}        
\end{figure}
In the simulations, muons starting from a source (Gaussian distribution with $\sigma=1$~mm, momentum of \SI{27}{MeV/c} and momentum spread \SI{0.8}{MeV/c}) traverse the first two layers, producing hits $p_1 = (x_1,y_1,z_1)$ in layer 1 and $p_2 = (x_2,y_2,z_2)$ in layer 2, and then stop in a sample; see Fig.~\ref{track_matching}. A stopped muon decays and emits a positron which traverses two detector layers and produces hits $p_3 = (x_3,y_3,z_3)$ and $p_4 = (x_4,y_4,z_4)$ (in any set of detectors). The emitted positron, of high mean kinetic energy $\Bar{E} = \SI{36.9}{MeV}$, experiences much smaller scattering compared to surface muons. 
Given the coordinates $p_1$ and $p_2$ for the muon, its track is extrapolated to the sample's $z$-position, $p_\mu = (x_\mu,y_\mu,z_s)$. Similarly, using $p_3$ and $p_4$ for the corresponding decay positron, it is extrapolated to the sample z-position, $p_e = (x_e,y_e,z_s)$. Here, the $z$-axis is given by the incoming muon beam direction [see Fig.~\ref{spect}]. The extrapolated tracks deviate from the actual landing / decay position of the muon $p_s = (x_s,y_s,z_s)$ by a distance of $\delta_\mu=p_s-p_\mu$ for the muon and $\delta_e=p_s-p_e$ for the positron; see Fig.~\ref{track_matching}. $p_s$ is determined using a non-interactive (void) reference volume placed in front of the sample. In order to approximate the real tracking scheme, a muon-positron pair must satisfy a matching condition, which requires that the distance between $p_e$ and $p_\mu$ to be smaller than a set value of $d_{match}$, usually $\sim\SI{1}{mm}$. The standard deviation (STD) of $\delta_\mu$ and $\delta_e$ of all valid events is used as a measure of the goodness of the spectrometer's tracking. The applied magnetic fields are simulated using the calculated field map of the GPS instrument\cite{gps} at PSI, which was scaled to achieve the desired magnitude at the sample position along the $z$-direction (parallel to the incoming muon beam direction). The performance of the spectrometer was investigated for magnetic fields with magnitudes between \SI{0}{T} and \SI{0.75}{T}, which correspond to the actual limits of the GPS magnet. In addition, the performance of the spectrometers with the spacing between the inner layers in the range of \SIrange{10}{40}{mm} is studied. 

\section{Results}
\subsection{Linear detector configuration}
We start by simulating the linear configuration with the field applied along the beam direction $z$. From these simulations, we find that the application of a field up to a value of \SI{\sim 50}{mT} has only a minimal effect on the tracking capabilities of the simulated spectrometer; see Fig.~\ref{linear_diff_d_diff_B}.
In particular, the STD value generally increases linearly with the magnitude of the applied field. This is true for both incoming muons and emitted positrons. However, since the muon trajectories are almost parallel to the applied field, the STD of $\delta_\mu$ exhibits a very small field dependence, i.e. a variation below 2\% in the range of \SIrange{0}{0.75}{T}. In contrast, the effect of the applied field on the positrons is more dramatic due to their almost randomly oriented trajectories relative to the applied field. In fact, for positrons, the slope of the STD vs. $d$ increases with the applied field. For fields larger than \SI{50}{mT}, the degradation in tracking capability becomes significant, as reflected in the increase in STD of $\delta_e$.

To better understand the effect of the magnetic field on the tracking quality, we plot the STD of $\delta_\mu$ and $\delta_e$ as a function of applied field at a fixed distance $d=\SI{20}{mm}$ in Fig.~\ref{b_vs_sigma}. This particular $d$ was chosen because it corresponds to the typical radius of a cryostat's radiation shield on GPS. For this spectrometer configuration, the application of a magnetic field up to \SI{\sim 50}{mT} results in an increase of less than 2\% compared to the STD values of $\delta_e$ and $\delta_\mu$ in  zero field. In higher fields, we observe an almost linear increase in the STD of $\delta_e$, reaching roughly twice the zero field value at the maximum applied field of \SI{0.75}{T}. Note that the effect of the field on the STD of $\delta_\mu$ is less than $2\%$ irrespective of the applied field.

\subsection{Cuboidal detector Configuration}
\begin{figure}[htb]
    \centering
    \includegraphics[width=0.45\linewidth]{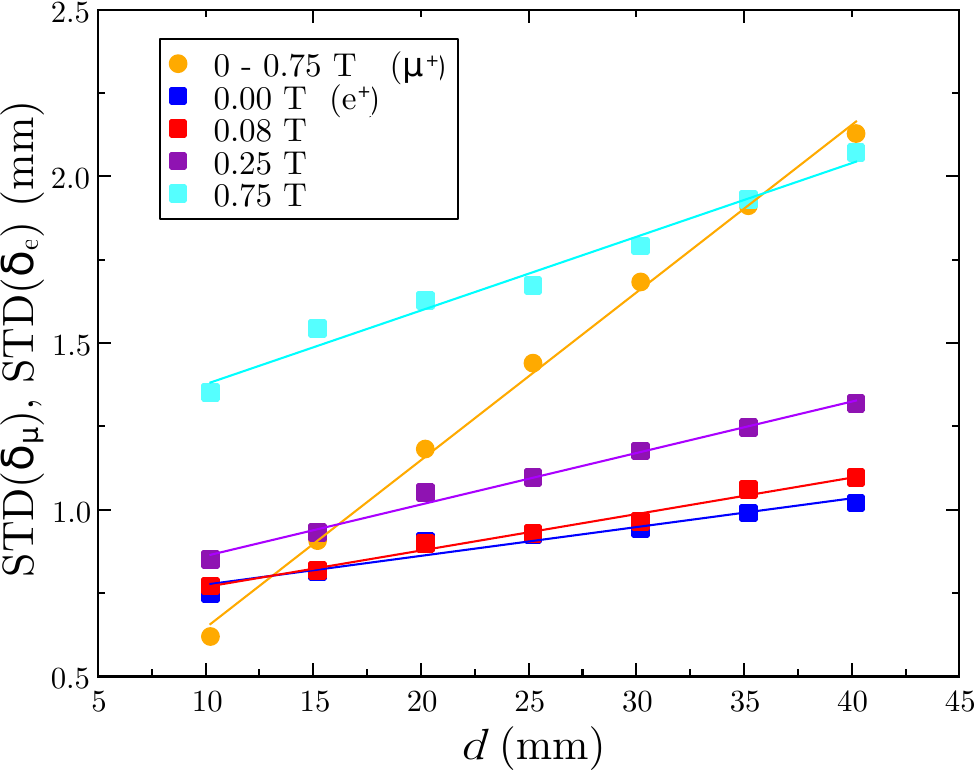}
    \caption{The STD of $\delta_\mu$ and $\delta_e$ in the cuboidal configuration as a function of the separation between the sample and the inner detector layers for different external magnetic field magnitudes in the range from \SIrange{0}{0.75}{T}. The STD of $\delta_\mu$ for different field magnitudes is represented by one data set, as the sets differ by less than 2\%.}
    \label{cuboid_diff_d_diff_B}
\end{figure}

Similarly to the linear configuration, the deflection of the muon trajectories due to the magnetic field is also minimal in the cuboidal configuration. The STD of $\delta_\mu$ increases by less than 2\% for fields up to $\SI{0.75}{T}$ compared to their zero field values; see Figs.~\ref{b_vs_sigma} and \ref{cuboid_diff_d_diff_B}. This is to be expected given that the overall geometry with regard to the incoming muons does not change significantly compared to the linear configuration. Although the STD of $\delta_e$ does not exhibit a strong field dependence below \SI{\sim 100}{mT}, at higher fields we observe an almost linear increase in the STD of $\delta_e$, as seen also in the linear configuration.
However, the slope of the STD vs. field for the cuboidal configuration is lower than that for the linear one. This can be explained by the fact that only positrons passing through the smaller inner layer and the larger outer layer are detected as a valid event. Therefore, curved trajectories, e.g. for lower-energy positrons, are not detected as valid events in this geometry since they likely miss one of the layers.

\section{Discussion}
In order to utilize the power of Si pixel-based spectrometers for \mSR\ applications, one should use the tracking information provided by the Si chips to produce vertex-reconstructed \mSR\ (vx-\mSR) spectra \cite{mandok2025arxiv}. As discussed earlier, an incoming muon is matched to its decay positron by requiring that its extrapolated landing position on the sample be within $d_{match}$ distance from the extrapolated origin of the positron at the sample's $z$ position. Experimentally, the positron must be detected within a $t_{gate}$ time after the muon arrives to the sample. However, since the simulations provide an event-by-event information, we do not consider this timing condition here.

\begin{figure}[htb]
    \centering
    \begin{minipage}[t]{0.45\textwidth}
    \centering
    \includegraphics[width=\linewidth]{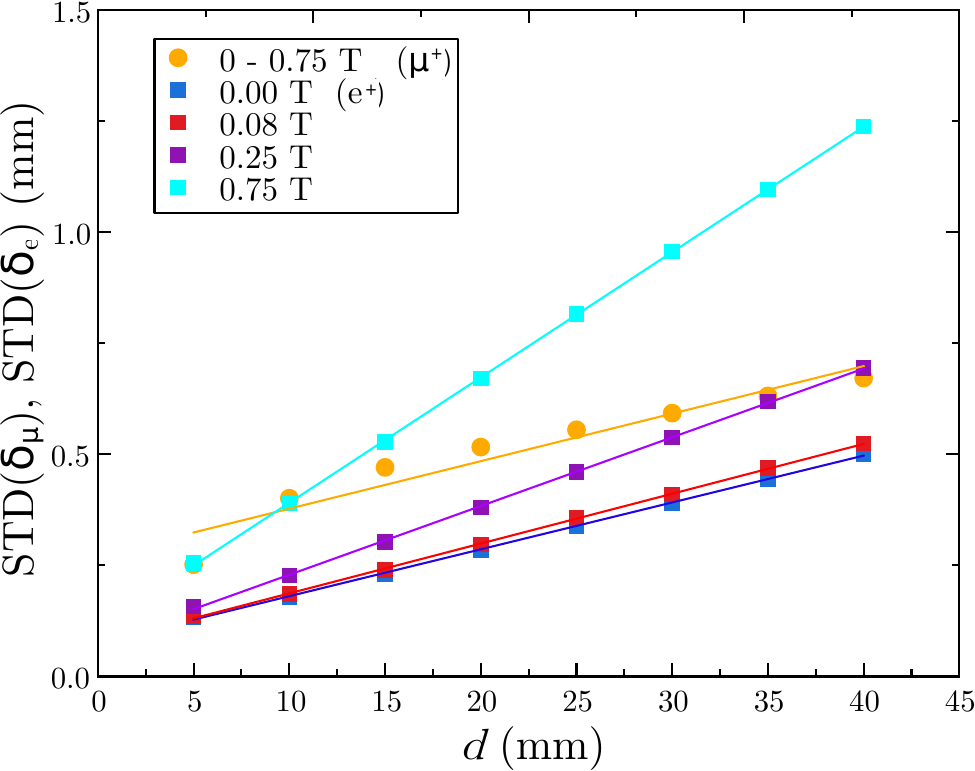}
    \caption{The STD of $\delta_\mu$ (circles) and $\delta_e$ (squares) for good events in the linear configuration as a function of the separation between the sample and the inner detector layers for different applied magnetic fields.}
    \label{fig:match-linear}
    \end{minipage}\hfill
    \begin{minipage}[t]{0.45\textwidth}
    \centering
    \includegraphics[width=\linewidth]{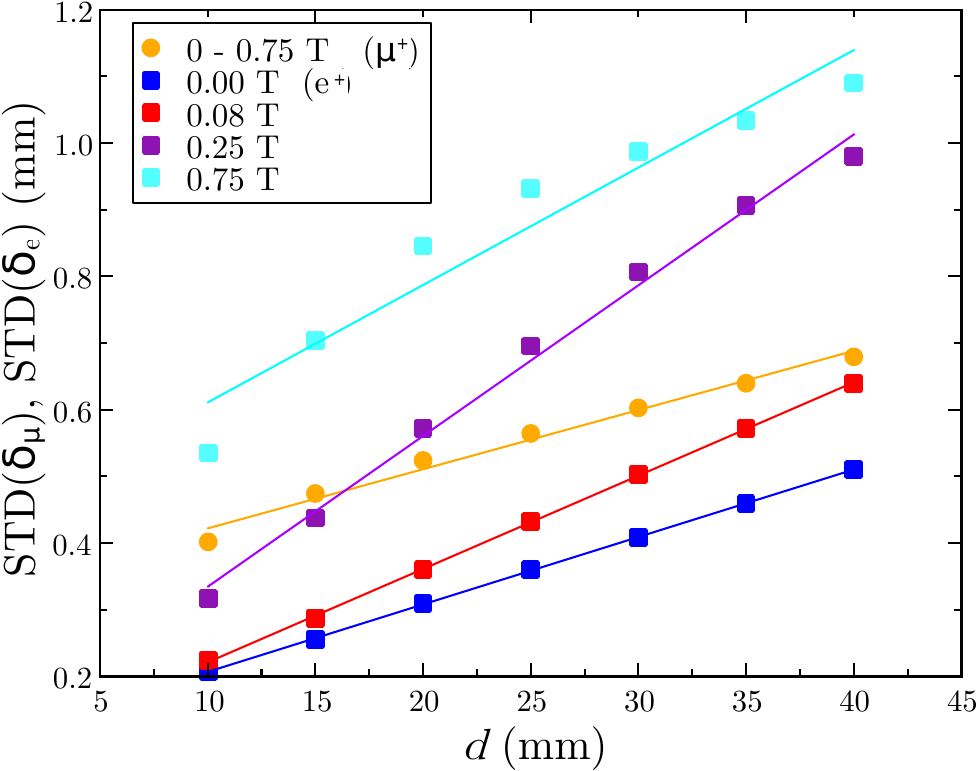}
    \caption{The STD of $\delta_\mu$ (circles) and $\delta_e$ (squares) for good events in the cuboidal configuration as a function of the separation between the sample and the inner detector layers for different applied magnetic fields.}
    \label{fig:match-cuboid}
    \end{minipage}
\end{figure}

Taking the muon and positron hit coordinates from the data in the previous section, we determine whether the extrapolated trajectories meet the matching criterion with $d_{match} = 1$~mm. Matching events are taken as good events, while all other events are discarded. This allows us to evaluate the performance and suitability of the different spectrometer configurations for vx-\mSR\ measurements in zero and applied magnetic fields. Figs. \ref{fig:match-linear} and \ref{fig:match-cuboid} show the STD values of $\delta_\mu$ and $\delta_e$ considering only good events for the linear and cuboidal configurations, respectively.

\begin{figure}[!h]
    \centering
    \begin{minipage}[t]{0.45\textwidth}
    \includegraphics[width=\linewidth]{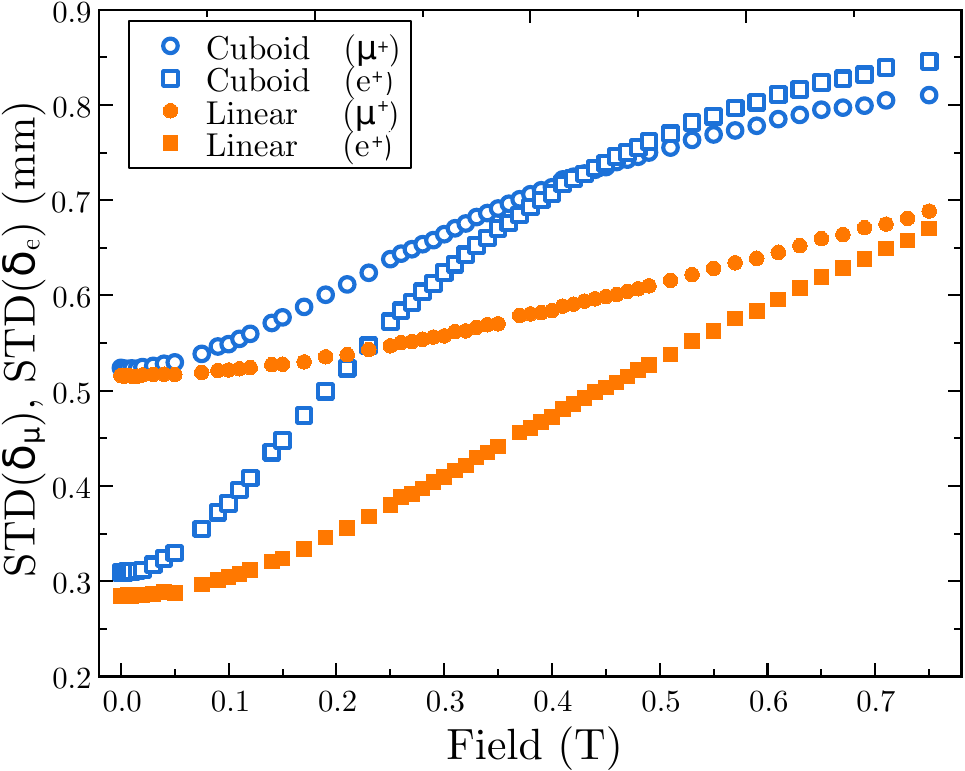}
    \caption{The STD of $\delta_\mu$ (circles) and $\delta_e$ (squares) for matched (good) events in the linear (orange) and the cuboidal (blue) configurations as a function of applied magnetic field for a fixed distance $d=20$~mm between the samples and inner layers.}
    \label{fig:MatchSTD}
    \end{minipage}\hfill
    \begin{minipage}[t]{0.45\textwidth}
    \centering
    \includegraphics[width=\linewidth]{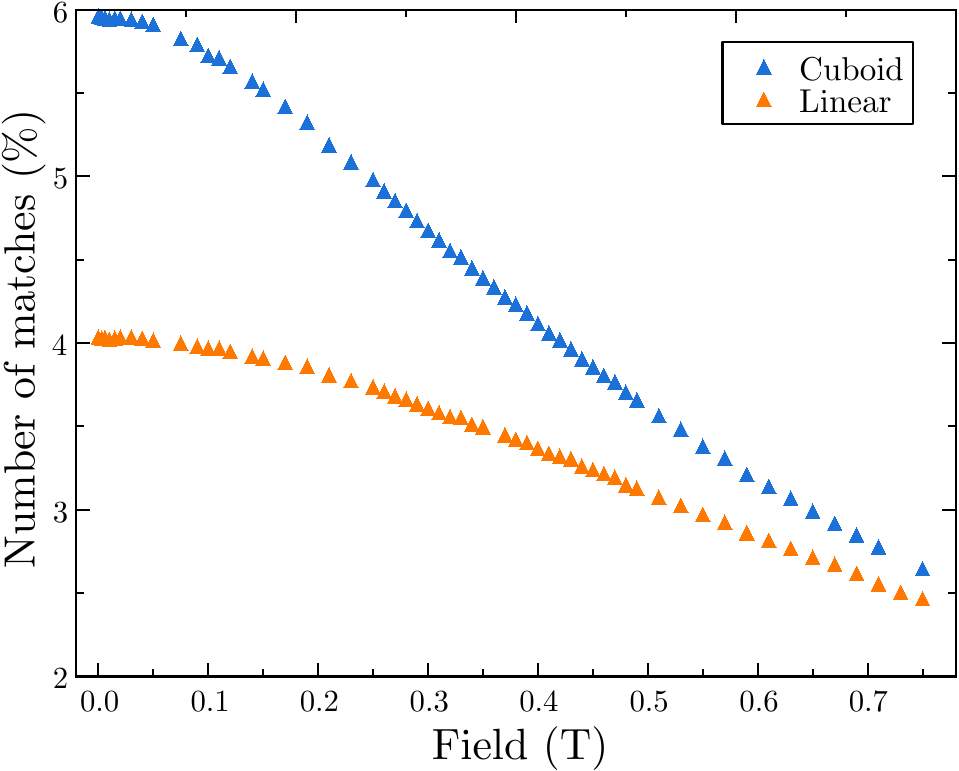}
    \caption{The fraction of matched (good) events from the total number of detected positrons as a function of applied field in both configurations.}
    \label{fig:MatchRatio}
    \end{minipage}
\end{figure}

Note that the linear dependence on $d$ is maintained in both cases (i.e. compared to Figs.~\ref{linear_diff_d_diff_B} and \ref{cuboid_diff_d_diff_B}). However, it is important to note that the STD values are significantly smaller in this case. This can be clearly seen in Fig.~\ref{fig:MatchSTD} where we plot the STD values for the matched events as a function of the applied field (compared to Fig.~\ref{b_vs_sigma}).
This can be understood by considering the stringent conditions imposed on these good events. This condition naturally selects less curved (due to the magnetic field) and less deflected (due to multiple scattering) trajectories, which results in a degradation in the counting rate. In Fig.~\ref{fig:MatchRatio} we plot the fraction of matched events from the total number of detected positrons. As can be seen, this fraction decreases gradually with increasing magnetic field primarily due to discarding more curved trajectories at higher fields.

\section{Summary and Conclusions}
We simulated the effects of applied magnetic fields on particle tracking capabilities in two different configurations of a \mSR\ spectrometer using Si pixel detectors. The linear configuration consisting of two upstream and two downstream detector layers is currently being tested at PSI \cite{mandok2025arxiv}. From these simulations, we find that the application of magnetic fields in the beam direction do not affect our ability to accurately track the landing position of incoming muons on the sample using only two Si pixel layers. However, the effect of the magnetic field on the emitted positrons is crucial. In fact, the simulations indicate that for a two-layered Si pixel detector spectrometer, one can reconstruct the trajectories of particles reasonably well in a magnetic field of up to \SI{\sim50}{mT}. For configurations that can incorporate a cryostat of a diameter of \SI{40}{mm}, similar to those used at PSI, we estimate that a lateral resolution of the order of \SI{\sim 1}{mm} is possible in this low magnetic field regime. For operation in higher magnetic fields, i.e. \SI{>50}{mT}, tracking of decay positrons may be possible only using a third Si pixel layer. Alternatively, one may devise a trajectory reconstruction scheme based on an accurate knowledge of the magnetic field map relative to the detector configuration and geometry.


\section{Acknowledgments}
This research is funded by the Swiss National Science Foundation (SNF-Grant No. 200021\_215167).


\bibliographystyle{iopart-num}
\bibliography{main}

\end{document}